\documentclass[final,3p]{CSP}

\usepackage{lineno}
\usepackage{amsmath}
\usepackage{comment}
\usepackage[acronyms,nomain,xindy,nowarn]{glossaries}
\usepackage{algorithm}
\usepackage{siunitx}
\usepackage{wasysym}
\usepackage{graphicx}
\usepackage[outdir=./]{epstopdf}
\usepackage{afterpage}
\usepackage{placeins}
\usepackage{subfigure}
\usepackage{pgfplots}
\usepgfplotslibrary{colorbrewer}
\usepackage{tikz}
\usetikzlibrary{arrows.meta}
\usepackage{multirow}
\usepackage{changepage}
\usepackage{algorithm}
\usepackage{algpseudocode}
\usepackage{caption}
\usepackage{float}

\newacronym{ask}{ASK}{amplitude-shift keying}
\newacronym[plural={AWGN}, \glsshortpluralkey={AWGN}]{awgn}{AWGN}{additive white Gaussian noise}

\newacronym{ber}{BER}{bit error rate}
\newacronym{bler}{BLER}{block error rate}
\newacronym{bpsk}{BPSK}{binary phase-shift keying}
\newacronym{bsc}{BSC}{binary symmetric channel}

\newacronym{cdf}{CDF}{cumulative distribution function}
\newacronym{cgh}{CGH}{computer generated hologram}
\newacronym{cnn}{CNN}{convolutional neural network}
\newacronym{csi}{CSI}{channel-state information}
\newacronym{csit}{CSI-T}{channel-state information at the transmitter}

\newacronym{dmd}{DMD}{digital micromirror device}
\newacronym{dsp}{DSP}{digital signal processing}
\newacronym{ds}{DS}{direct search}

\newacronym{ecc}{ECC}{error-correcting code}

\newacronym{fde}{FDE}{frequency domain equalisation}
\newacronym{ff}{FF}{feed-forward}
\newacronym{ffnn}{FF-NN}{feed-forward neural network}
\newacronym{fsk}{FSK}{frequency-shift keying}

\newacronym{gm}{GM}{Gaussian mixture}
\newacronym{gd}{GD}{group delay}
\newacronym{gi}{GI}{graded-index}

\newacronym{hg}{HG}{Hermite-Gaussian}

\newacronym{ic}{IC}{interference channel}
\newacronym{iid}{i.\,i.\,d.}{independent and identically distributed}
\newacronym{il}{IL}{insertion loss}
\newacronym{iot}{IoT}{Internet-of-Things}

\newacronym{lb}{LB}{lower bound}
\newacronym{ldpc}{LDPC}{low density parity check}
\newacronym{lp}{LP}{linearly polarized}
\newacronym{lg}{LG}{Laguerre-Gaussian}

\newacronym{mac}{MAC}{multiple access channel}
\newacronym{map}{MAP}{maximum a posteriori}
\newacronym{mc}{MC}{mode coupling}
\newacronym{lmc}{LMC}{linear mode coupling}
\newacronym{mer}{MER}{mode extinction ratio}

\newacronym{mdl}{MDL}{mode-dependent loss}
\newacronym{mimo}{MIMO}{multiple-input multiple-output}
\newacronym{mmf}{MMF}{multimode fibre}
\newacronym{ml}{ML}{machine learning}
\newacronym{mop}{MOP}{multi-objective programming problem}
\newacronym{mplc}{MPLC}{multi-plane light conversion}
\newacronym{mrc}{MRC}{maximum ratio combining}
\newacronym{mse}{MSE}{mean squared error}

\newacronym{nn}{NN}{neural network}
\newacronym{nve}{NVE}{normalized validation error}

\newacronym{pdf}{PDF}{probability density function}
\newacronym{pwc}{PWC}{polar wiretap code}
\newacronym{pls}{PLS}{physical layer security}
\newacronym{pl}{PL}{photonic lantern}

\newacronym{qam}{QAM}{quadrature amplitude modulation}
\newacronym{qkd}{QKD}{quantum key distribution}

\newacronym{rbf}{RBF}{radial basis function}
\newacronym{relu}{ReLU}{rectified linear unit}
\newacronym{ris}{RIS}{reconfigurable intelligent surface}
\newacronym{rnn}{RNN}{recurrent neural network}

\newacronym{sc}{SC}{selection combining}
\newacronym{sdm}{SDM}{space-division multiplexing}
\newacronym{sgd}{SGD}{stochastic gradient descent}
\newacronym{siso}{SISO}{single-input-single-output}
\newacronym{sgp}{SGP}{secure goodput}
\newacronym{sinr}{SINR}{signal-to-interference-plus-noise ratio}
\newacronym{slm}{SLM}{spatial light modulator}
\newacronym{smf}{SMF}{single-mode fibre}
\newacronym{snr}{SNR}{signal-to-noise ratio}
\newacronym{svd}{SVD}{singular value decomposition}
\newacronym{svm}{SVM}{support vector machine}

\newacronym{tin}{TIN}{treating interference as noise}
\newacronym{tvd}{TVD}{total variation distance}
\newacronym{tm}{TM}{transmission matrix}

\newacronym{ub}{UB}{upper bound}
\newacronym{urllc}{URLLC}{ultra-reliable low latency communication}

\newacronym{wma}{WMA}{wavefront matching algorithm}

\newacronym{xt}{XT}{modal crosstalk}

\newacronym{zoc}{ZOC}{zero-outage capacity}
\setacronymstyle{long-short}

\newcommand{\todo}[2][]{
	\if\relax\detokenize{#1}\relax
	{\color{red}[TODO: #2]}%
	\else
	{\color{red}[TODO (#1): #2]}%
	\fi
}

\begin{document}

\begin{frontmatter}

\title{Unlocking Mode Programming with Multi-Plane Light Conversion Using Computer-Generated Hologram Optimisation}

\author[1,5]{Stefan~Rothe}
\author[2]{Fabio~Barbosa}
\author[3,4,6]{Jürgen~W.~Czarske}
\author[2,7]{Filipe~M.~Ferreira}

\address[1]{Department of Applied Physics, Yale University, New Haven, CT 06520, USA}
\address[2]{Optical Networks, Dept Electronic \& Electrical Eng, University College London, UK}
\address[3]{TU Dresden, Faculty of Electrical and Computer Engineering, Laboratory of Measurement and Sensor System Technique, 01062 Dresden, Germany}
\address[4]{TU Dresden, Faculty of Physics, 01062 Dresden, Germany}
\address[5]{stefan.rothe@yale.edu}
\address[6]{juergen.czarske@tu-dresden.de}
\address[7]{f.ferreira@ucl.ac.uk}





\begin{abstract}\rm
\begin{adjustwidth}{2cm}{2cm}{\itshape\textbf{Abstract:}} 
Programmable optical devices provide performance enhancement and flexibility to spatial multiplexing systems enabling transmission of tributaries in high-order eigenmodes of spatially-diverse transmission media, like \gls{mmf}. Wavefront shaping with \glspl{slm} facilitates scalability of the transmission media by allowing for channel diagonalization and quasi-single-input single-output operation. 
Programmable mode multiplexing configurations like \gls{mplc} utilise the SLM and offer the potential to simultaneously launch an arbitrary subset of spatial tributaries in any $N$-mode MMF. Such programmable optical processor would enable the throughput of \gls{sdm} systems to be progressively increased by addressing a growing number of tributaries over one \gls{mmf} and in this way meet a growing traffic demand -- similarly to the wavelength-division multiplexing evolution path. Conventionally, \gls{mplc} phasemasks are calculated using the \gls{wma}. 
However, this method does not exploit the full potential of programmable mode multiplexers. We show, that computer-generated hologram algorithms like direct search enable significant improvement 
compared to the traditional \gls{wma}-approach. Such gains are enabled by tailored cost functions with dynamic constraints concerning insertion loss as well as mode extinction ratio. We show that average mode extinction ratio can be greatly improved by as much as $\approx \SI{15}{\decibel}$ at the expense of insertion loss deterioration of $< \SI{3}{\decibel}$.
One particular feature of programmable mode multiplexers is the adaptability to optimised transmission functions. 
Besides conventional LP modes transmission, we employ our approach on Schmidt modes, which are spatial eigenchannels with minimum crosstalk derived from a measured transmission matrix.
\end{adjustwidth}
\end{abstract}

\end{frontmatter}

\section{Introduction}
\Gls{sdm} is an established approach to overcome the capacity exhaustion in optical single-mode fibre networks that underlie the nonlinear Shannon-limit~\cite{essiambre2010capacity,richardson2010filling,richardson2013space,puttnam2021space}. Amongst \gls{sdm} implementation options, the approach based on \glspl{mmf} allows the highest information density and consequently the highest efficiency gains regarding array integration in transceivers and other network elements. Usually, undesired effects in \gls{sdm} operation with \gls{mmf} such as group delay 
spread~\cite{arik2016group}, and crosstalk~\cite{ferreira2017semi} are compensated on the receiver-side via \gls{mimo}-\gls{dsp}~\cite{ryf2012mode,mori2014few}. Other effects such as \gls{mdl}~\cite{ho2011mode} require advanced detection algorithms~\cite{vblast}. {Furthermore, the group delay spread and the \gls{mdl} loss have been mitigated through the cyclic mode-group permutation technique \cite{cyclic} making all information tributaries to have similar spreading reducing the memory depth of \gls{mimo}-\gls{dsp}.}


The linear crosstalk 
in \gls{mmf} is due to linear mode coupling, 
refractive-index inhomogenities or small deviations of the core-cladding boundary caused by perturbations introduced during the fabrication process or by mechanical stresses imposed on the \gls{mmf} field. These imperfections cause the \gls{mmf} modes to couple among each other. As such, when exciting a pure mode at the \gls{mmf} input, some of its power is scattered to other guided modes. This power transfer results in a spreading of the channel impulse response due to characteristic group velocity~\cite{ferreira2017semi}. Therefore, the equalisation of the received signal must span over a time window that covers all the significant distortions undergone by a given information symbol. The ultimate objective of this process is to end up with the lowest possible crosstalk after equalization, or in positive terms, the highest possible \gls{mer}. 

The complexity required on the receiver-side increases significantly with the number of modes used. Despite the number of information tributaries $M$ actually transmitted, all $N$ fibre modes need to be detected such that \gls{mimo}-\gls{dsp} can be performed, otherwise \gls{mdl} would lead to frequent outage in the presence of non-negligible linear crosstalk. 
This leads to a complexity increase not only in terms of computational complexity (where \gls{mimo} complexity scales with $N^2$), but also in terms of hardware with the number of coherent receiver front-ends required being equal to the number of fibre modes $N$~\cite{InanComplexity}.  
While \glspl{mmf} provide great potential for spatial multiplexing efficiency gains, with as much as $N=1000$ modes~\cite{Filipe1000}, is what makes this particularly critical, since it would not be feasible to have $N$ coherent front-ends at system beginning-of-life but only at system end-of-life.

Here, we propose a programmable optical processor enabling scaling the number of information tributaries over an optical fibre with $N>M$ to progressively exploit the fibre spatial domain --
similar to the successful development of wavelength division multiplex systems with multiple wavelength channels as traffic demand grows. 
Using channel knowledge through, for example, the optical transmission matrix~\cite{popoff2011controlling,choi2011transmission,vcivzmar2012exploiting,carpenter2016complete,xiong2018complete,mazur2019characterization,rothe2019transmission,Cheng2023,zhong2023,cao2023controlling} and a programmable mode multiplexer, one can find the required wavefront that compensates for crosstalk, for a defined sub-domain of spatial modes. In order to transmit such tailored signals to the receiver with high fidelity, precise modulation of the amplitude, and especially the phase of the input wavefront for each desired spatial mode is necessary. {Here, we neglect any polarisation mixing in the \gls{mmf}.}

Photonic lanterns are fused-fibre devices that map a 1D spot array (or another arrangement of non-overlapping spots) provided by a collection of \gls{smf} to a 2D superposition of spatial modes interfacing an \gls{mmf}~\cite{leon2010photonic}. The underlying approach is to form a multimode waveguide by gradually transition the \gls{smf} array/bundle into a composite multimode waveguide~\cite{fontaine2022photonic}. There are several fabrication processes that all aim to conserve entropy allowing for a nearly lossless transition. {As a result, photonic lanterns provide perhaps the highest optical coupling efficiency of usually less than 1~dB \gls{il}, which makes it particularly promising for photon sensitive applications such as quantum networks~\cite{alarcon2021few,alarcon2023dynamic}}. However, the photonic lantern suffers from limitations in its reconfigurability. Although there are solutions reporting a mode-selective transmission of a few, e.g. 6 modes over a comparably short \gls{mmf} length of $\SI{100}{\meter}$~\cite{velazquez2015six}, the \textit{programmability} is largely limited before, and even impossible after the photonic lantern has been fabricated. In particular, mitigation of degenerate mode coupling turns out to represent the major hurdle and can hardly be achieved despite clever \gls{smf} arrangement or microstructured preforms for large mode numbers above 15~\cite{Fontaine12,Velazquez18}). This limits the achievable selectivity to the mode-group-level.

{
The \gls{slm} is probably the device that allows for the highest flexibility for modulating the amplitude and phase of a wavefront. 
Given an \gls{slm} and an appropriate wavefront shaping algorithm on the transmitter-side, any tailored signal, i.e. any arbitrary mode combination can be provided to the receiver~\cite{vellekoop2015feedback,cao2022shaping,gomes2022near}.
For instance Schmidt modes~\cite{ho2013linear}, also referred to as transmission eigenchannels~\cite{choi2011transmission,mosk2012controlling,kim2012maximal,sarma2016control,yilmaz2019transverse,cao2022shaping}, could be launched and detected in order to pre- and post-compensate, respectively, MMF-induced crosstalk and thus allow  (partial) channel diagonalisation. Alternatively, 
principal modes~\cite{fabioPMs,carpenter2015observation} suffer minimal mode dispersion and distortion. These specific mode sets enable improvement on information density with scalable \gls{dsp} equalisation complexity~\cite{fabioPMsJournal}. Moreover, mode combinations can be extracted from the transmission matrix generating a confidential data link to a legitimate recipient~\cite{rothe2023securing}.} Thus, information security, \gls{dsp} complexity and spatial multiplexing gain can be scaled simultaneously and flexibly with an \gls{slm}. The key to high performance is to find the right wavefront building a high-order mode multiplexer transmitting the desired mode combinations associated to $M$ respective spatial sub-channels. 

{
One flexible and high-dimensional mode multiplexing approach which can be realised with an \gls{slm} is \gls{mplc}~\cite{labroille2014efficient,fontaine2017design,fontaine2019laguerre,mounaix2020time,fontaine2021hermite, rademacher2021peta,wen2021scalable,fontaine2022photonic,kupianskyi2023high,pohle2023intelligent}. Moreover, \gls{mplc} setups are also used for coherent beam combining~\cite{billaud2019optimal}, untangling light that outputs a \gls{mmf}~\cite{kupianskyi2024all}, quantum cryptography~\cite{fickler2020full}, computational imaging~\cite{butaite2022build}, and diffractive neural networks / optical computing~\cite{lin2018all,zhou2022photonic,yildirim2024nonlinear}. }
The \gls{mplc} apparatus is an all-optical device that enables mapping of spatially disjoint input spots to spatially overlapping mode fields. 
In most \gls{sdm} applications that employ an \gls{mplc}, the phasemasks are calculated \textit{a priori} as part of an optimisation procedure based on \gls{wma}~\cite{sakamaki2007new}. The results are displayed on the \gls{slm} or etched on a reflective surface performing the desired transformation. 
In that way, the spatial bandwidth of \glspl{mmf} can be exploited gainfully achieving data rates in the $\SI{}{\peta\bit\per\second}$ range via only one single core of standard \gls{mmf}~\cite{rademacher2021peta}. 

In the context of a communication system, however, the \gls{wma} does not provide a global optimum for average \gls{mer}, i.e. $\overline{\mathrm{MER}}$, and simultaneously minimum \gls{il}. 
{
Recent experimental work on multiplexing 250 \gls{hg} modes in a graded-index MMF shows \gls{il} ranging from -3.5~dB to -2.5~dB, and crosstalk ranging from -6.8~dB to -2.1~dB~\cite{wen2021scalable}.
Even though there are significant experimental breakthroughs involved in these results, the achieved fidelity of the multiplexer cannot be regarded as sufficient to generate specific mode sets that would be required for channel diagonalisation in long MMF transmission links covering hundreds of km.
}
In this paper, we show that by using tailored cost functions, the \gls{mplc} fidelity can be significantly improved using computer-generated hologram algorithms~\cite{fickler2020full,tsang2018review,rothe2021benchmarking}. 
We demonstrate that by scaling the cost function constraints, the degree of optimisation, namely maximum $\overline{\mathrm{MER}}$ and / or minimum \gls{il}, can be tuned flexibly. 
{
In our simulations, we find that the further optimisation on $\overline{\mathrm{MER}}$ leads to additional reduction of the \gls{mdl}. In recent work, flexible cost function constraints were employed within a gradient ascent optimisation framework~\cite{kupianskyi2023high}. Here, our approach integrates the traditional \gls{wma} into a direct-search environment, leveraging a different optimisation setting.}
We define the solution space of our approach with respect to the number of tributaries $M$ using a fixed \gls{slm} pixel number. Our results enable customised optimisation for high-performance \gls{sdm} communication systems with high number of utilised spatial channels transmitting \gls{lp} modes, as well as Schmidt modes. 
{
This is particularly significant, since recent works focus on shaping \gls{hg} modes to explore the inherent symmetry of Canonical SLMs and allowing only for mode group selectivity in MMF. In turn, cylindrically symmetric LP modes that are the MMF eigenmodes can be directly addressed by our approach. 
Therefore, our work contributes to efficient and arbitrary wavefront shaping of cylindrical mode sets using an \gls{mplc} device.
}

\section{Parameter selection and starting point}
\label{sec:parameters}

\begin{figure}[t]
\centering
(a)\includegraphics[height=5cm]{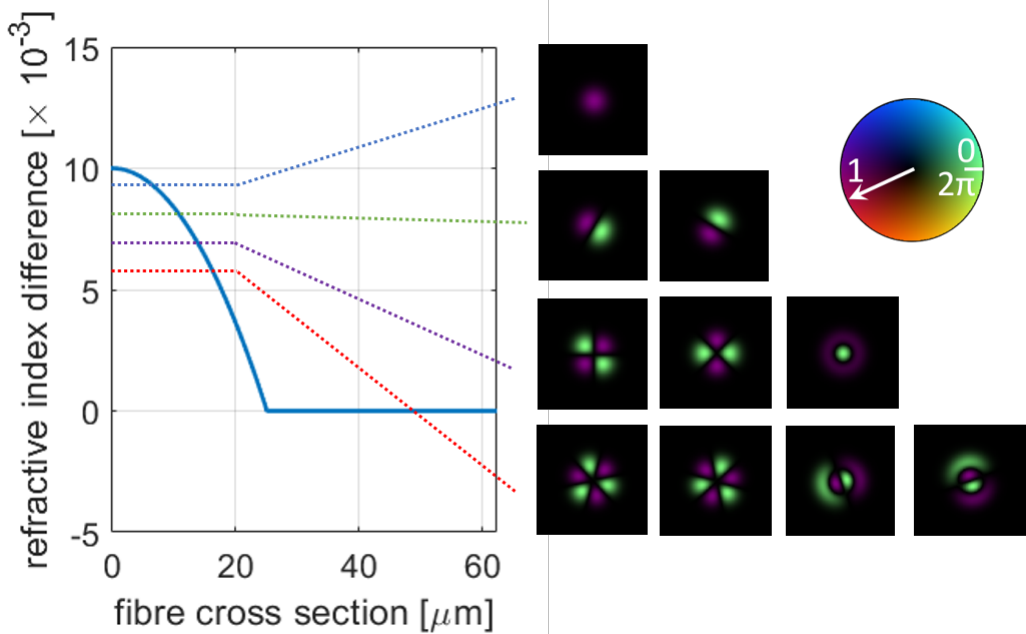}
(b) \includegraphics[height=5cm]{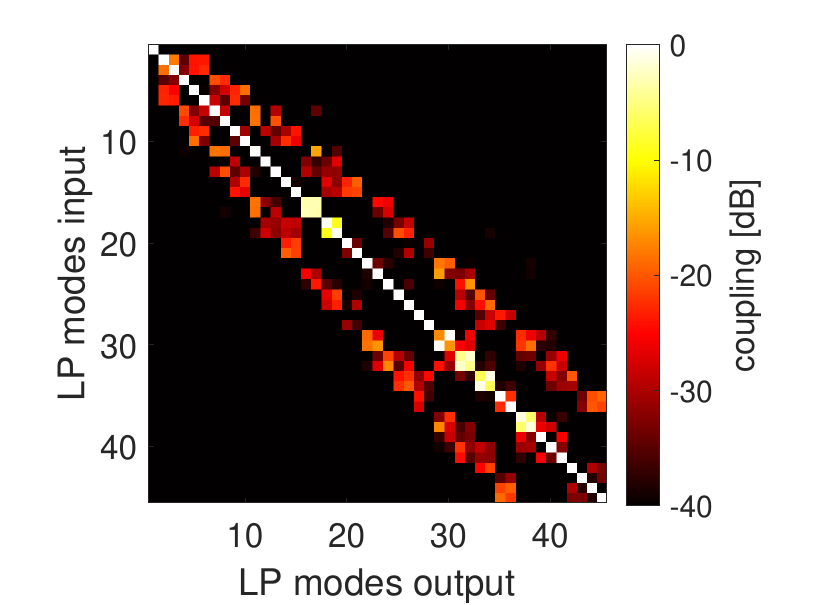}\\
(c) \includegraphics[height=5cm]{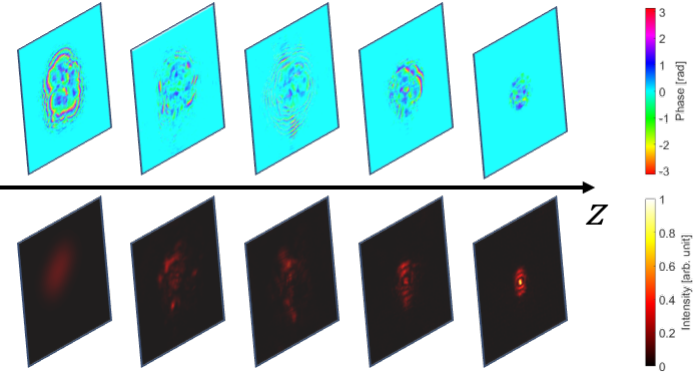}   
\caption{ (a) Refractive index profile of the considered \gls{mmf}. This \gls{mmf} supports $N=45$ \gls{lp} modes per polarisation, of which the first 10 modes corresponding to the first 4 lowest order-groups are shown on the right-hand side. (b) Measured transmission matrix of $\SI{10}{\kilo\meter}$ \gls{mmf} using photonic lantern as mode (de-) multiplex and coherent detection. (c) Phasemasks~(top) and intensity profiles~(bottom) of a simulated $K=5$-passage \gls{mplc} for $M=5$ tributaries. We have randomly drawn the target modes from the 45-mode set of the considered MMF. Disjoint Gaussian input spots transition into an incoherent superposition of \gls{lp} modes at the output by using \gls{wma}. We normalised phase (intensity) plots to the same phase (intensity) scale. Colorbars for phase and intensity are shown on the right. We used the code provided in Refs.~\cite{fontaine2019laguerre,fontaine2019laguerreSupp} to produce this result.}
\label{fig:fig1}
\end{figure}

We assume a standard $\diameter \SI{50}{\micro\meter}$-core \gls{mmf} with graded-index: core grading exponent equal to $1.97$ and core relative refractive index difference equal to $0.01$. Such fibre guides $45$ \gls{lp} modes per polarisation at $\SI{1550}{\nano\meter}$ wavelength. The fibre refractive index profile is shown 
in Fig.~\ref{fig:fig1}a.
Although the fabrication of this optical fibre is highly precise, crosstalk is inevitably present after long transmission distance. 
Usually, \gls{mmf} exhibits a minimum of $\SI{-20}{\decibel}$ crosstalk after $\SI{100}{\kilo\meter}$ transmission. 
This phenomenon can be traced in Fig.~\ref{fig:fig1}b that shows the transmission matrix after $\SI{10}{\kilo\meter}$ transmission, where off-diagonal entries indicate the presence of crosstalk. 
For this measurement, we used photonic lantern-based mode (de-)multiplex and standard coherent detection. 
Most methods compensating crosstalk underlie powerful \gls{dsp} paradigms applied on the receiver-side, whose computational effort increases exponentially 
with each added tributary. 
To allow flexibility gains in \gls{sdm} systems, however, the number of tributaries should refer to a scalable parameter without causing adverse performance impact. 
Here, we introduce transmitter-side precompensation using a programmable mode multiplexer, i.e. wavefront shaping with an \gls{slm}. 
Following the \gls{mplc} approach reported in Ref.~\cite{fontaine2019laguerre}, the \gls{slm} is being illuminated by $\SI{120}{\micro\meter}$ spaced input beams that are arranged in an rectangular array with Gaussian profile and $\SI{60}{\micro\meter}$ mode field diameter. Depending on the number of actually used tributaries $M$, the number of active input beams varies accordingly. 
We consider a fixed number of $K=5$ reflection passages (see discussion in Sec.~\ref{sec:wma}) each of which is modeled by simulating a $256\times256$ pixel grid with $\SI{8}{\micro\meter}$ pixels. The passages are connected with a $\SI{21}{\milli\meter}$ free-space propagation kernel. The entire apparatus is modelled in transmission. This being said, effects from input angle misalignment or shadowing of the beam path are neglected.
Fig.~\ref{fig:fig1}c shows exemplary phasemasks for a $K=5$-passage configuration and intensity profiles at the corresponding axial position along the propagation direction $z$~(indicated with a black arrow). The phasemasks shown are the result of our multiplexer design and carry out the desired transformation from $M=5$ disjoint input beams into the incoherent superposition of 5 output modes. 

Within this work, we present a mode multiplexer that achieves greatly improved performance on the aforementioned wavefront shaping task. 
In this context, there are two crucial parameters we use as cost for optimisation. First, the excitation of each target mode should be as precise as possible. In particular, the output of the mode multiplexer should exhibit as high purity as possible that is the coupling to a target mode calculated as complex spatial overlap integral. At the same time, the overlap to residual modes should be as low as possible achieving minimum crosstalk.
In order to quantify the achievable purity of the mode multiplexer, the \gls{mer} is determined calculating  the ratio between the power in the $m$th mode and the power in all other $N$ modes (i.e. reciprocal representation of crosstalk)

\begin{equation}
    \mathrm{MER}_m= 20 \cdot \mathrm{log}_{10} \left( \frac{ \left\vert \int \int E_{\textrm{m}}^{*}E_{\textrm{m}}
     \right\vert}
    {
    \sum_{i\neq m}^N{
    \left\vert 
     \int \int E_{m}^* E_{i} \right \vert
     }
    }
    \right )
    ,~\overline{\mathrm{MER}}=\frac{1}{M}\sum_m^M\mathrm{MER}_m
    \label{eq:mer}
\end{equation}

\noindent where $E_m^*$ is the conjugate mode multiplexer output field in the target \gls{lp} mode and $E_i$ indicates the field of any other \gls{lp} mode. 
As such, performance on coupling efficiency for all $M$ tributaries can be determined by calculating the average $\overline{\mathrm{MER}}$.
Second, it is desirable to carry out an power-efficient shaping process. Therefore, we calculate the \gls{il} that provides information on how much light is lost from input to output of the multiplexer. We calculate \gls{il} by the average singular value $s_v$ of the multiplexer's coupling matrix

\begin{equation}
    \mathrm{IL}=20 \cdot \mathrm{log}_{10} \left(
    \overline{s_v}
    \right),
    \label{eq:il}
\end{equation}

\noindent where the operation $\overline{ ( \cdot )  }$ indicates averaging over all singular values $s_v$~\cite{fontaine2019laguerre}. 
{When transmitting data over spatially parallel paths, equal power distribution, i.e. low mode dependency, is desirable. Therefore, in our simulations we characterise \gls{mdl}$=10 \cdot \mathrm{log}_{10} \left( s_{v,\text{max}}^2/s_{v,\text{min}}^2 \right)$~\cite{wen2021scalable}, yet do not implement it as cost function for optimisation.}

\section{Implementation of wavefront matching and direct search optimisation}
\label{sec:wma}

\subsection{ Wavefront matching}
Mode multiplexing for a high number of tributaries has become a pivotal desire. Over the past decade, the \gls{mplc} approach has become subject of intense research activities.
Depending on both the input spot distribution and target mode profiles, phasemasks are required that enable the desired beam transformation. 
Usually, the \gls{wma} is used to precalculate such phasemasks and it works as follows~\cite{sakamaki2007new}. 
{
We assume propagation along the optical axis $z$, and define both an input field distribution at the input plane, as well as a target field distribution at the output plane. In between, there are $K$ equally spaced planes at which the forward propagating beam can experience spatial phase modulation. We refer to these planes as (reflection) passages, as in a real experiment an \gls{slm} would induce phase modulation in a reflection configuration.  
In traditional \gls{mplc} setups, the \gls{wma} is used to calculate the spatial overlap at each passage between forward propagating input field, e.g. an arrangement of disjoint Gaussian spots
, and output target modes propagating in backward direction. 
The phase conjugate mismatch between spots and modes at each passage is computed and displayed as phase modulation in the following iteration.
After a certain number of iteration passes, this algorithm converges to a smooth transition from input field to target output field.}

{
It has been shown that \gls{wma} accomplishes high multiplexing throughput with a comparably low number of reflection passages. 
Such gains in the required number of passages are achieved by harnessing Canonical symmetries in rectangular mode sets such as \gls{hg} modes that align well on the rectangular pixel grid of an \gls{slm}~\cite{fontaine2017design,fontaine2019laguerre,fontaine2021hermite}. 
However, these advantages are specific to systems designed to utilize \gls{hg} modes.}

{
Here, the focus of our work is on generating \gls{lp} modes. In rotationally symmetric fibres, \gls{lp} modes form complete orthogonal set under the scalar approximation, making them the appropriate basis for describing signal propagation~\cite{gloge1971weakly,snyder1983optical,cao2023controlling}. 
Specifically, for fibres with a parabolic refractive index profile—such as graded-index fibres, the modes are best represented using Laguerre-Gaussian polynomials, often referred to as LG modes.
For this reason, \gls{lp} modes are usually used in experimental demonstrations on high-throughput \gls{sdm} communications~\cite{rademacher2021peta}.}

{
In turn, \gls{hg} modes with Canonical symmetries are not inherently suited for rotationally symmetric fibres, or spherical resonator configurations~\cite{siegman1973hermite}, as they do not satisfy the boundary conditions required for cylindrically symmetric fibres.
Instead, \gls{hg} modes are more appropriate for rectangular waveguides or free-space propagation, where their boundary conditions are naturally satisfied.} 

{
Our work therefore focuses on the more generally deployed set of \gls{lp} modes, as well as Schmidt modes (e.g. also used for step-index fibers), which are usually more difficult to shape with the \gls{wma} due to the lack of Canonical symmetries.
As such, the generation of \gls{lp} modes is associated with a dramatic increase in the number of required \gls{slm} passages. While recent work on \gls{lp} mode transmission shows that for each spatial tributary, one passage is required, the limited number of \gls{slm} pixels is no longer sufficient.} 
Therefore, in many applications the phasemasks are rather printed on a dielectric mirror, which leads to near-perfect reflection properties, but makes flexibility after fabrication impossible.
Here, we restrict the number of reflection passages to as much as $K=5$ and thereby obtain results that are implementable on a real \gls{slm} with limited pixel number. Therefore, we can utilise programmable precompensation approaches using e.g. spatial eigenchannels, as described in Sec.~\ref{sec:results}. 

\begin{figure}[t]
    \centering
    \begin{tabular} {c c}
        \multirow[b]{2}{*}[2.6cm]{(a)\includegraphics[width=0.49\textwidth]{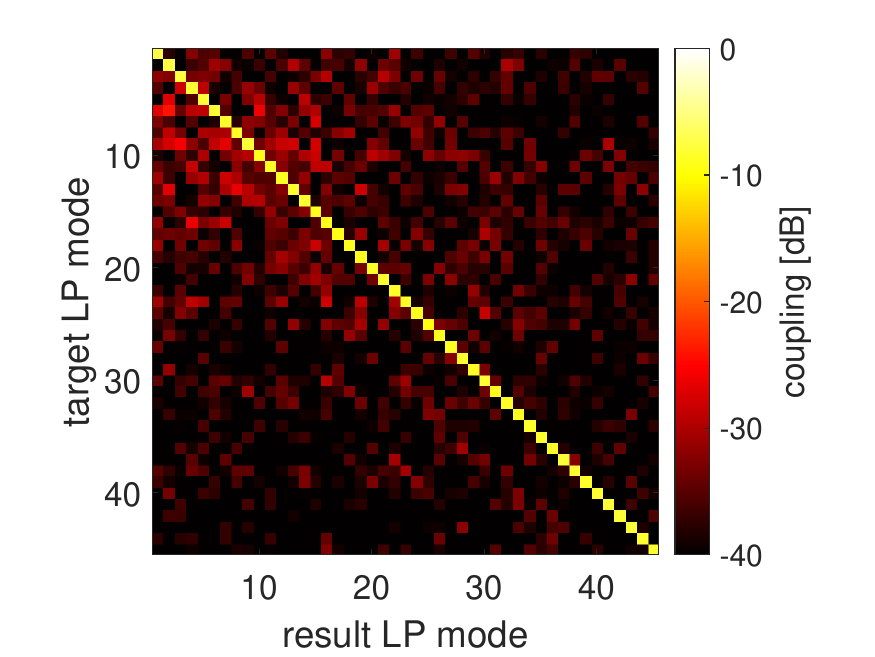}} & (b)\includegraphics[width=0.37\textwidth]{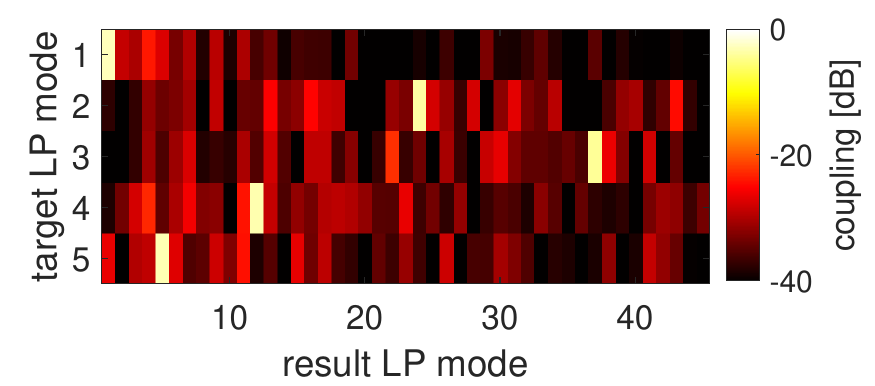}\\
        & (c)\includegraphics[width=0.37\textwidth]{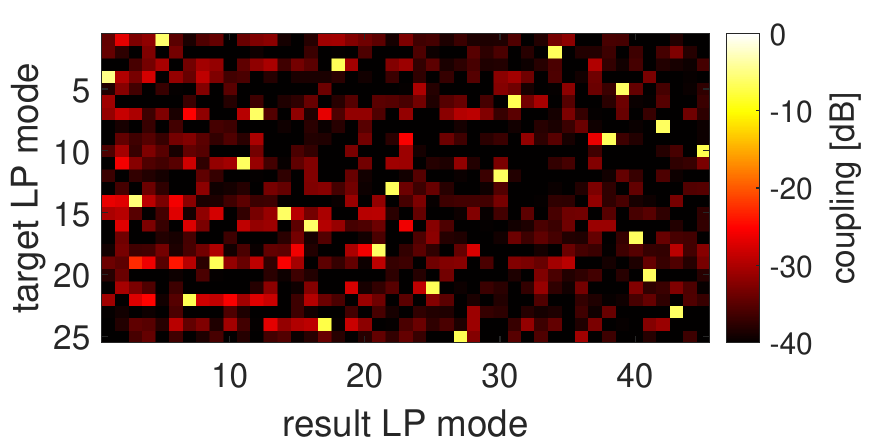}\\
    \end{tabular}
    (d)\includegraphics[height=5cm]{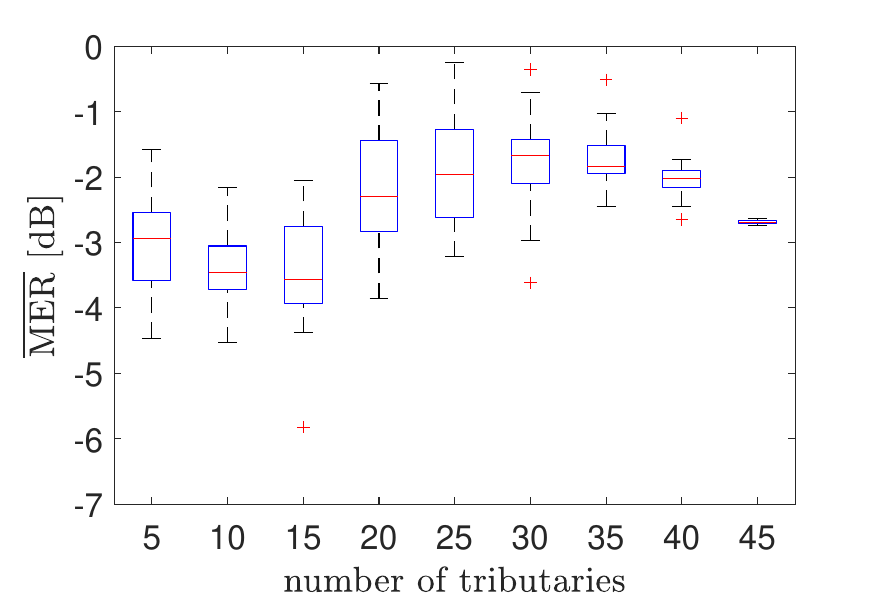}
    (e)\includegraphics[height=5cm]{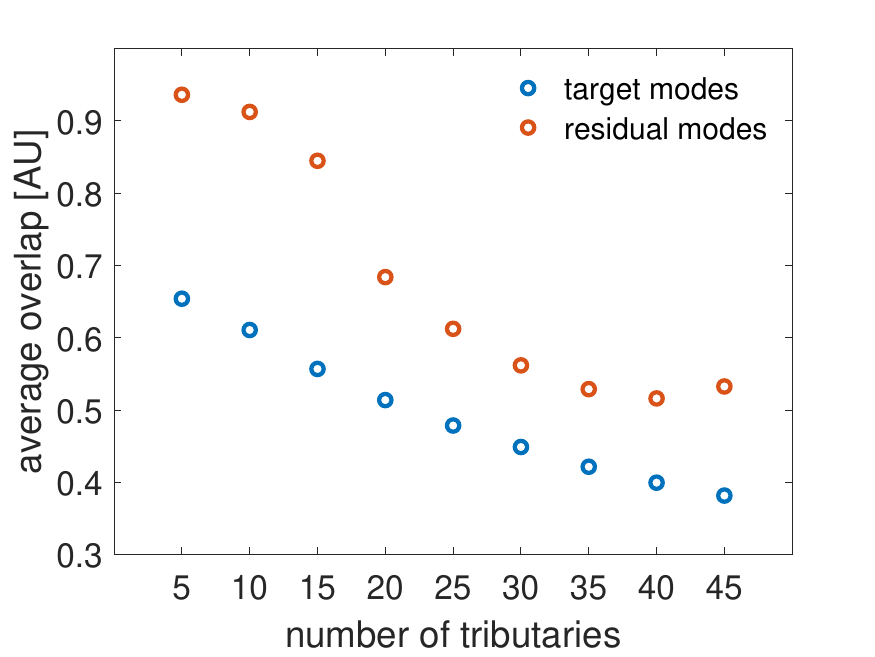}
    \caption{(a) \gls{wma} result when shaping all 45 \gls{lp} modes the \gls{mmf} supports ($M=N$). (b) \gls{wma} result when shaping a random subset of $M=5$ tributaries out of $N=45$ \gls{lp} modes. (c) Same principle as (b), but with a $M=25$-tributary subset. (d) average \gls{mer} box-plotted over different tributary numbers. Each boxplot corresponds to 30 different randomised subsets of the defined number of tributaries. (e) Average overlap to both target and residual modes under varying number of tributaries.}  
    \label{fig:wma}
\end{figure}

Fig.~\ref{fig:wma}a shows the \gls{wma} result for multiplexing all 45 \gls{lp} modes the \gls{mmf} mentioned in Sec.~\ref{sec:parameters} supports, that is $M=N$, using 5 reflection passages. The amplitude of the device's coupling matrix is shown. Note, that ideal coupling equals a coupling factor of 1, i.e. $\SI{0}{\decibel}$. In this realisation, an average \gls{mer} of $\SI{-4.25}{\decibel}$ and \gls{il}$=\SI{-8.28}{\decibel}$ results. 
Considerable crosstalk of the mode multiplexer is clearly present, which inevitably leads to performance penalty in an \gls{sdm} system, if no mitigating steps are taken. 
Apparently, paralleled shaping of 45 modes in such a highly restricted setup is simply too demanding for a 5-passsage \gls{mplc} and one could expect improvement if the requirements are relaxed.

Let us use less tributaries (i.e. $M<N$) and shape an arbitrary subset of target modes using the \gls{wma}. The result for shaping only 5 randomly selected tributaries is shown in Fig.~\ref{fig:wma}b, while the equivalent for 25 tributaries is shown in Fig.~\ref{fig:wma}c. For the 5-tributary case, the average \gls{mer} equals $\SI{-2.78}{\decibel}$ and \gls{il}$=\SI{-3.07}{\decibel}$. In the case of 25 tributaries, an average \gls{mer} of $\SI{-2.17}{\decibel}$ and \gls{il}$=\SI{-6.07}{\decibel}$ is achieved. 
In Fig~\ref{fig:wma}d, boxplots indicate average \gls{mer} (i.e. $\overline{\mathrm{MER}}$) values under varying number of tributaries. For each data point, we produced 30 \gls{wma} runs in each of which a random triutary subset with fixed $M$ is shaped. Examples for the 5 and 25-tributary case are shown in Fig.~\ref{fig:wma}b and Fig.~\ref{fig:wma}c, respectively. It turns out that simply relaxing the shaping effort by reducing the number of tributaries does not provide a universal remedy for high-performance multiplexing using the \gls{wma}.    
The underlying reason for this is the \gls{wma} principle, which regulates only the overlap between forward propagating spots and backward propagating target modes. 
What the \gls{wma} does not consider, however, is suppression of coupling to residual modes reducing unwanted crosstalk.
Therefore, with a reduced tributary number it is more likely that deviations of the \gls{wma} will lead to overlap with a mode other than the target modes.
At constant number of degrees of freedom (i.e. \gls{slm} pixels), the overlap fidelity to target modes does enhance with decreasing number of tributaries, since the ratio between number of \gls{slm} pixels and number of modes increases. Fig.~\ref{fig:wma}e (blue curve) shows the average coupling to target modes. But at the same time, undesired coupling to residual modes also enhances. Fig.~\ref{fig:wma}e (red curve) shows the total coupling to residual modes. Providing a sanity check, the curves in Fig.~\ref{fig:wma}e allow to determine the optimal number of tributaries in terms of \gls{mer}. Since the target curve is always below the curve indicating coupling to residual modes, the \gls{mer} is best where the distance between the two curves is smallest. This is the case for about 30 tributaries, which is in line with Fig.~\ref{fig:wma}d.

\subsection{Dynamic cost function design using direct search}
In order to increase the mode multiplexer performance, we propose a novel cost function design tailored to \gls{mplc}. 
We incorporate the wavefront shaping task into a \gls{ds} framework~\cite{seldowitz1987synthesis,clark1996direct,christopher2020holographic}. Using $K$ phasemasks $P(x,y,K)$ resulting from the \gls{wma} as initial guess, we randomly select a single phase element $p_i$ of one of the passages and allocate a random phase $p_r$. After propagating all $M$ input spots $E_{\mathrm{in},m}(x,y), m=1,2,...,M$ to the final passage, we calculate $\overline{\mathrm{MER}}$, \gls{il} as well as the standard devitiation $\sigma_{\mathrm{MER}}$. The phase change will be accepted in the $i$th iteration, if the cost function is satisfied. 

When including $\overline{\mathrm{MER}}$ to the cost, the algorithm does improve the ratio between coupling to target and coupling to residual modes, but it is likely that this goal is achieved by simply increasing the overall loss.
Therefore, we also consider \gls{il} in the cost function constraining the optimisation algorithm tightly, but also allow deliberate releases. 
Additionally, we define the maximum standard deviation $\sigma_{\mathrm{MER}}$ among all tributaries as $\SI{1}{\decibel}$.
This means, one phase change is accepted, if the $\overline{\mathrm{MER}}$ is improved, the \gls{il} is still above the lower bound and the standard deviation $\sigma_{\mathrm{MER}}$ is below $\SI{1}{\decibel}$.
Our \gls{ds} procedure is summarised in Algorithm~\ref{alg:alg}. Additionally, we provide a code example of our method in a repository that is publicly accessible and can be found under Ref.~\cite{RotheGithub}.

\begin{algorithm}
\caption{Our \gls{ds} algorithm}\label{alg:alg}
\begin{algorithmic}[1]
\State Initialise input beams $E_{\mathrm{in},m}(x,y)$ \Comment{ $m=1,2,...,M$}
\State Initialise  phasemasks $P(x,y,K)$ using the WMA as for instance in Ref.~\cite{fontaine2019laguerre} \Comment{ $K$ $\equiv$ number of passages}
\State Define termination criterion $\overline{\mathrm{MER}}_T$ and lower bound $\mathrm{IL}_{lb}$ 
\State Set $\overline{\mathrm{MER}}_i = -\infty~\mathrm{dB}$ 
\While{$\overline{\mathrm{MER}}_i \leq \overline{\mathrm{MER}}_T$} \Comment{ $i$ $\equiv$ current iteration}
\State Randomly select one phase element $p_i=P(x_r,y_r,K_r)$
\State $P(x_r,y_r,K_r) \gets p_r=2\pi \cdot \mathrm{rand()}$
\State propagate all beams $E_{\mathrm{in},m}(x,y)$ to last passage
\State calculate $\overline{\mathrm{MER}}_i$ as in (\ref{eq:mer}) and $\mathrm{IL}_i$ as in (\ref{eq:il}), and $\sigma_{\mathrm{MER}_i}$
\If{$\overline{\mathrm{MER}}_i > \overline{\mathrm{MER}}_{i-1}$ and $\mathrm{IL}_i > \mathrm{IL}_{lb}$ and $\sigma_{\mathrm{MER}_i} < 1~\mathrm{dB}$   }
    \State $P(x_r,y_r,K_r) = p_r$
\Else
    \State $P(x_r,y_r,K_r) = p_i$
\EndIf
\EndWhile
\end{algorithmic}
\end{algorithm}

\begin{figure}[H]
    \centering
    (a)\includegraphics[height=6.2cm]{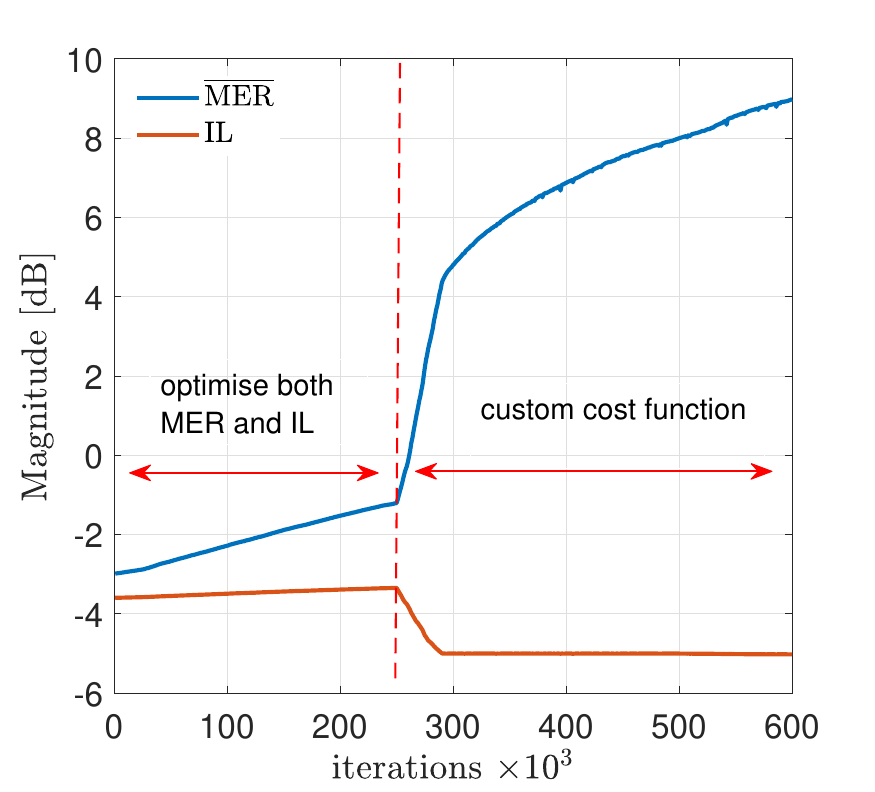}
    (b)\includegraphics[height=6.2cm]{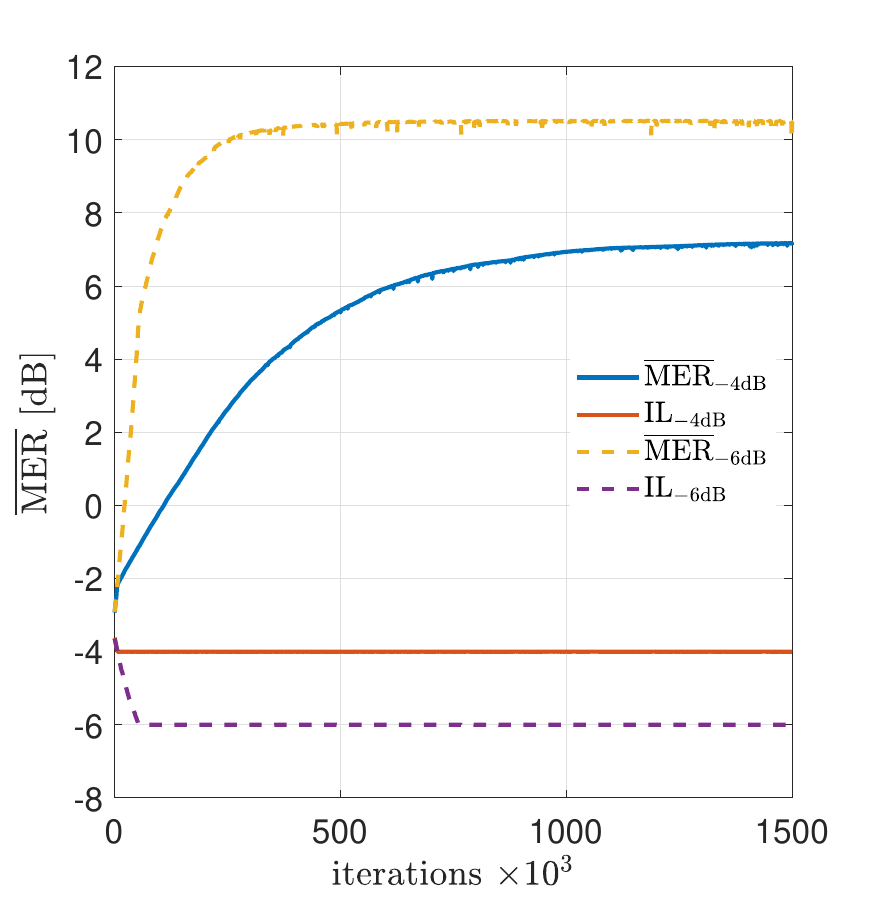}
    \\
    {(c)}\includegraphics[height=3.4cm]{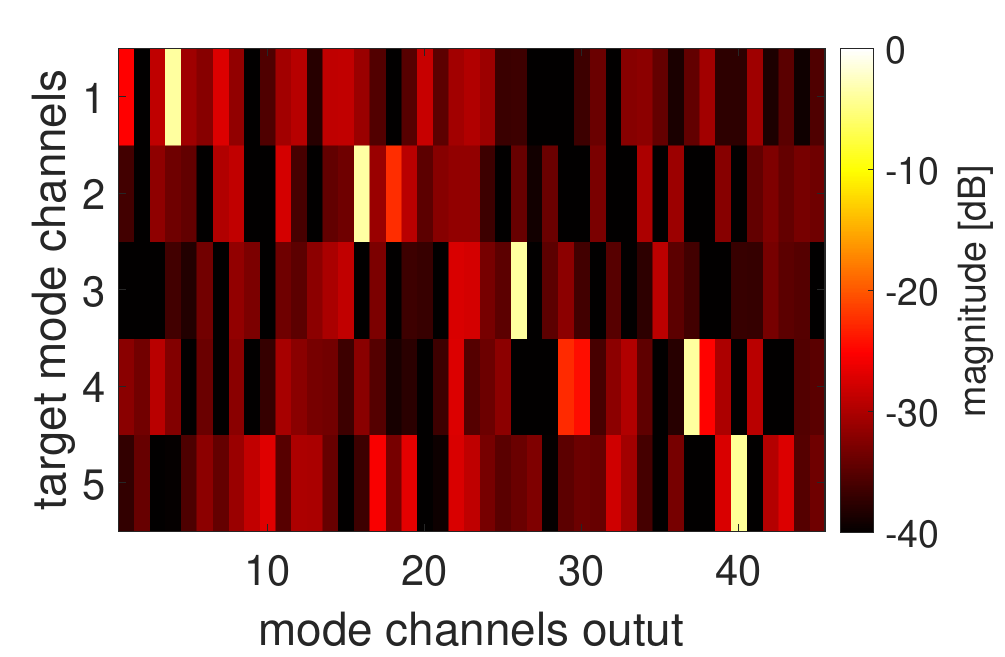}
    {(d)}\includegraphics[height=3.4cm]{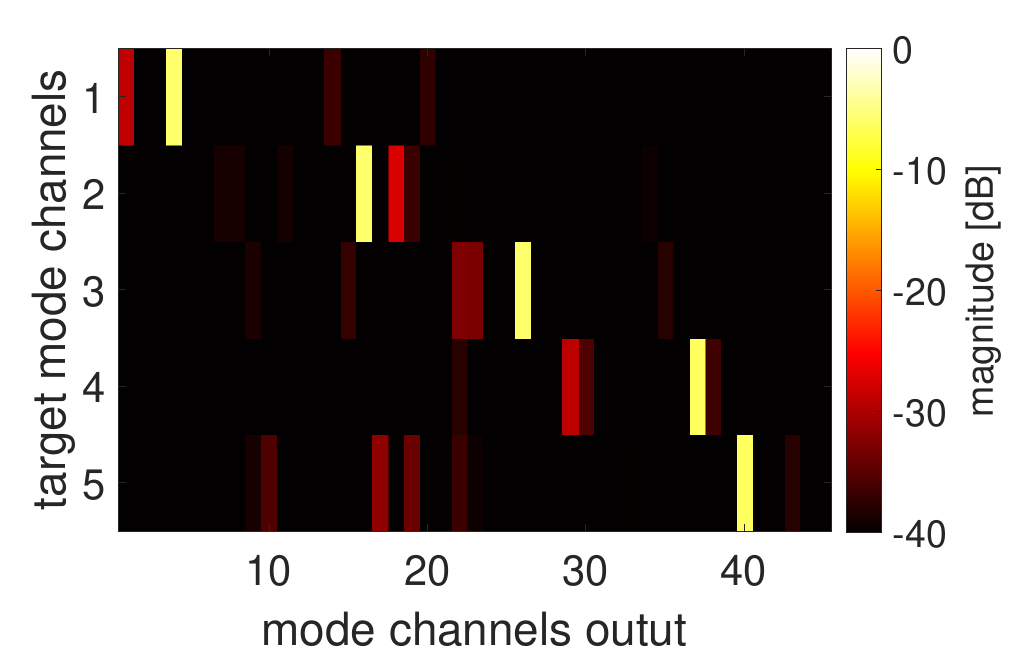}
    {(e)}\includegraphics[height=3.4cm]{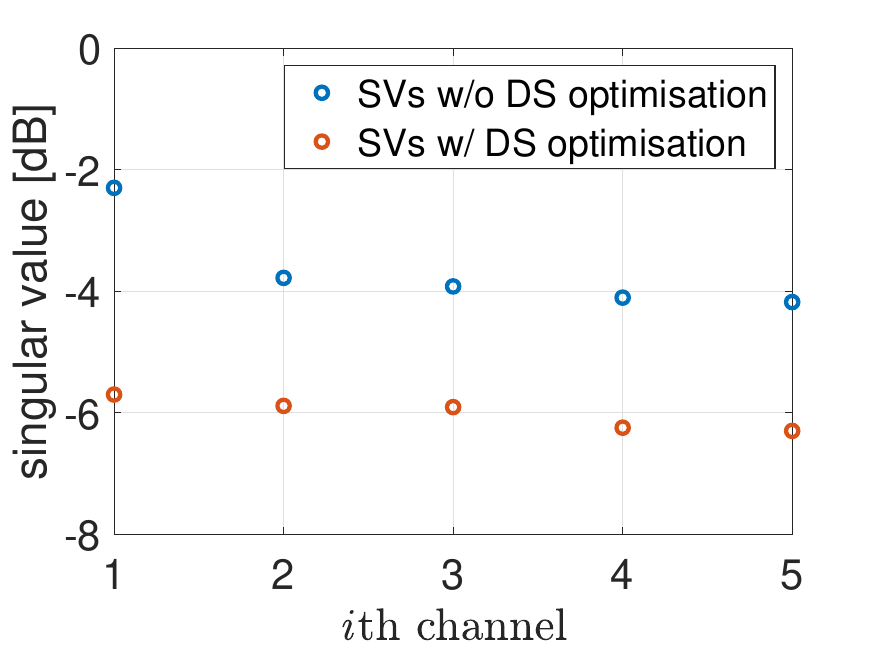}
    \caption{Improving mode multiplexing using \gls{ds}. When shaping $M=5$ out of $N=45$ \gls{lp} modes using the \gls{wma}, we get  $\overline{\mathrm{MER}}=-2.98$~dB and \gls{il}=-3.59~dB. (a) Now, we use the \gls{wma} result as initial guess for \gls{ds} and put both performance indicators into the cost function. If we try to reduce cost on both $\overline{\mathrm{MER}}$ and \gls{il}, the gains are negligible after 250k iterations. The average \gls{mer} has improved only by 1.7~dB, while the \gls{il} improved only 0.2~dB. However, if we deliberately relax the constraints and allow \gls{il} deterioration to the minimum of -5~dB, this has a huge impact on $\overline{\mathrm{MER}}$, which improves by 9.2~dB within 250k iterations. Another relaxation to \gls{il}=-6~dB improves $\overline{\mathrm{MER}}$ for additional 1~dB within another 100k iterations. 
    (b) Now, we run the algorithm with a constant cost function avoiding local minima. We define a lower bound on \gls{il} of as much as -4~dB (blue and orange solid curves), the $\overline{\mathrm{MER}}$ converges at 7.1~dB after 1.5 million iterations and improves by one order of magnitude in total. 
    If we lower the \gls{il} threshold to -6~dB (yellow and purple dashed curve), $\overline{\mathrm{MER}}$ improves by as much as 13.5~dB. {(c) and (d) show the coupling matrix of the multiplexer before (c) and after (d) \gls{ds} optimisation corresponding to a lower IL bound of -6~dB (yellow and purple curves in (b) ). We find strong suppression of coupling to unwanted modes. (e) Associated singular values of the multiplexer before and after DS optimisation. As targeted, the DS optimisation improves the $\overline{\mathrm{MER}}$ by slightly lowering the IL, i.e. the average singular value. This also lowers the variations among the singular values, that is MDL. In our case, e.g. for the yellow/purple curves in (b) (lower IL bound set to -6~dB), we automatically improve MDL from 0.94~dB calculated for the initial state to 0.29~dB after optimisation.}}  
    \label{fig:merIL5LPmodes}
\end{figure}

Figure~\ref{fig:merIL5LPmodes} illustrates an introductory example when using $M=5$ tributaries out of $N=45$ \gls{lp} modes. It shows the progression of both \gls{mer} and \gls{il} plotted over iterations.
The initial state provided by the \gls{wma} result is characterised with as much as $\overline{\mathrm{MER}}=\SI{-2.98}{\decibel}$ and \gls{il}$=\SI{-3.59}{\decibel}$ (compare Fig.~\ref{fig:wma}d, 5 tributaries).
Within the first 250k iterations, the algorithm seeks phase values to reduce cost on both $\overline{\mathrm{MER}}$ and \gls{il}.

We observe the gains are vanishing small, since neither $\overline{\mathrm{MER}}$ nor \gls{il} improve significantly.
While $\overline{\mathrm{MER}}$ has improved by $\SI{1.7}{\decibel}$, the \gls{il} improved only $\SI{0.2}{\decibel}$.
First, the initial state, that is the \gls{wma} result, is already good, yet not ideal. Second, with a limited set of degrees of freedom (i.e. number of \gls{slm} pixels) it is impossible to profit on all frontiers simultaneously. Note that a better hologram resolution, i.e. higher number of smaller pixels, could lead to overall improvement, but we restrict our study to the real-world case, where in most cases $\SI{8}{\micro\meter}$ \gls{slm} pixels are used.

Now, let us allow an \gls{il} deterioration and define a lower bound of $\SI{-5}{\decibel}$ corresponding to a \gls{il} penalty of $\SI{-1.6}{\decibel}$. 
After a few iterations, the algorithm relaxes the \gls{il} to the new lower threshold. 
At the same time, this has great impact on the $\overline{\mathrm{MER}}$ progression which increases by $\SI{9.2}{\decibel}$. 
If we lower the \gls{il} threshold even further to $\SI{-6} {\decibel}$, the \gls{il} does not exhaust completely, but only relaxes a little.
This behaviour indicates, that after running optimisation with specified constraints, the algorithm plateaus in a local minimum. 
In spite this fact, however, the $\overline{\mathrm{MER}}$ level is raised by another $\SI{1} {\decibel}$. 
In total, this direct search algorithm with dynamic cost function accomplishes an $\overline{\mathrm{MER}}$ gain by more than one order of magnitude with about $\SI{10.2} {\decibel}$. 
In exchange, an \gls{il} penalty of $\approx \SI{1.6} {\decibel}$ is tolerated. 

Fig.~\ref{fig:merIL5LPmodes}b shows the case, where we avoid intermediate plateaus and run the algorithm with only one, customised cost function. Allowing a small \gls{il} deterioration of $\SI{-0.4}{\decibel}$ (i.e. lower bound of  \gls{il} of $\SI{-4}{\decibel}$, blue and orange solid curves), the $\overline{\mathrm{MER}}$ reaches $\SI{2.6}{\decibel}$ after 250k. Compare this with Fig.~\ref{fig:merIL5LPmodes}a, where the $\overline{\mathrm{MER}}$ has reached $\SI{-1.2}{\decibel}$ after the same number of iterations. In that case, we started with the same initial state, but tried to optimise on both \gls{il} $\overline{\mathrm{MER}}$ and achieved an $\overline{\mathrm{MER}}$ improvement of half the slope compared to Fig.~\ref{fig:merIL5LPmodes}b.
In total, with slight \gls{il} deterioration and a relaxed cost function constraint, the $\overline{\mathrm{MER}}$ converges at $\SI{7.1}{\decibel}$ after 1.5 million iterations and improves by one order of magnitude. If we lower the \gls{il} threshold to even $\SI{-6}{\decibel}$ (yellow and purple dashed curve), $\overline{\mathrm{MER}}$ improves by as much as $\SI{13.5}{\decibel}$. 
{
Figures~\ref{fig:merIL5LPmodes}c and d show the two coupling matrices corresponding to the case where we set the lower IL bound to -6~dB, before we start our optimisation (c), and after we terminated it (d). When comparing the two results, we can clearly identify a strong suppression to unwanted modes which proves the improvement in $\overline{\mathrm{MER}}$. Figure~\ref{fig:merIL5LPmodes}e shows the singular values for the same case --- before and after DS optimisation. We can find that the average singular value, i.e. IL, slightly drops to -6~dB, as initiated.
Interestingly, in our simulations we automatically improve on variations among the singular values, i.e. \gls{mdl}. For instance, in the case where we set the lower IL bound to -6~dB, our optimisation improves MDL from 0.94~dB to 0.29~dB.}


\section{Transmission of Schmidt modes using direct search}
\label{sec:results}

Now we design the same \gls{mplc} mode multiplexer using a specialised mode basis and use it for an optimised data transmission. 
What becomes evident here is the extreme potential of the \gls{slm}. Thanks to its programmability, an enormous degree of flexibility is available. It allows to switch the mode base without the need to physically exchange a device, but easily load new phase masks onto its display.

\begin{figure}[H]
    \centering
    (a) \includegraphics[width=0.85\textwidth]{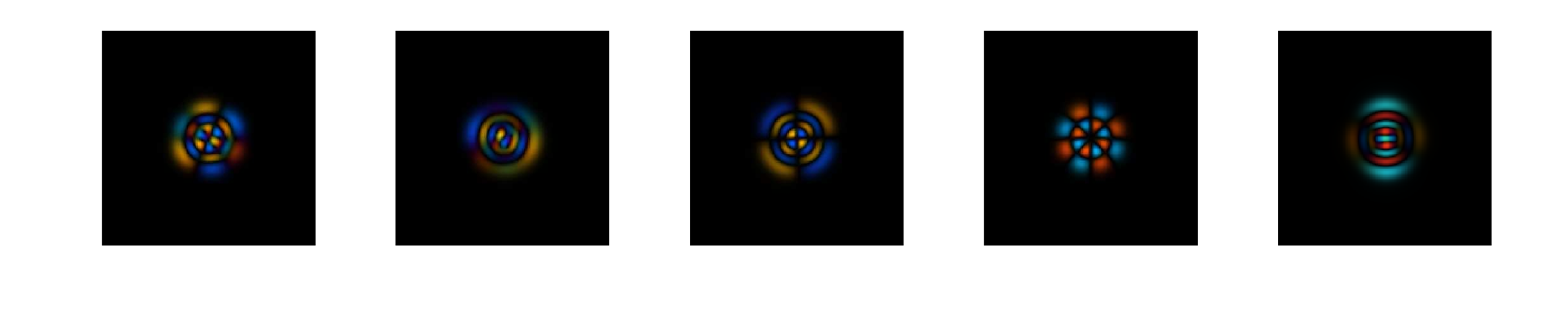} 
    \begin{minipage}{0.1\textwidth}
        \centering
        \raisebox{3.5cm}{\includegraphics[width=1.3cm]{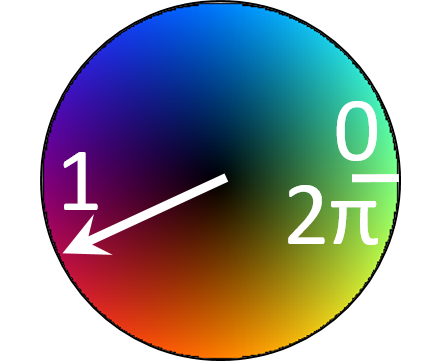}}\end{minipage}\\
        \vspace{-2.2cm}
    (b) \includegraphics[height=5.2cm]{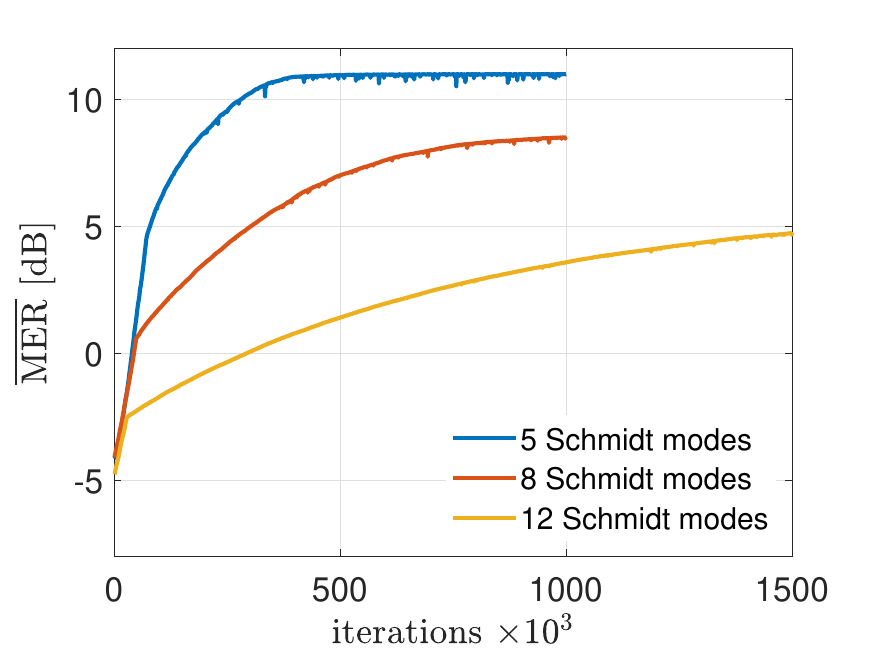} 
    (c)\includegraphics[height=5.2cm]{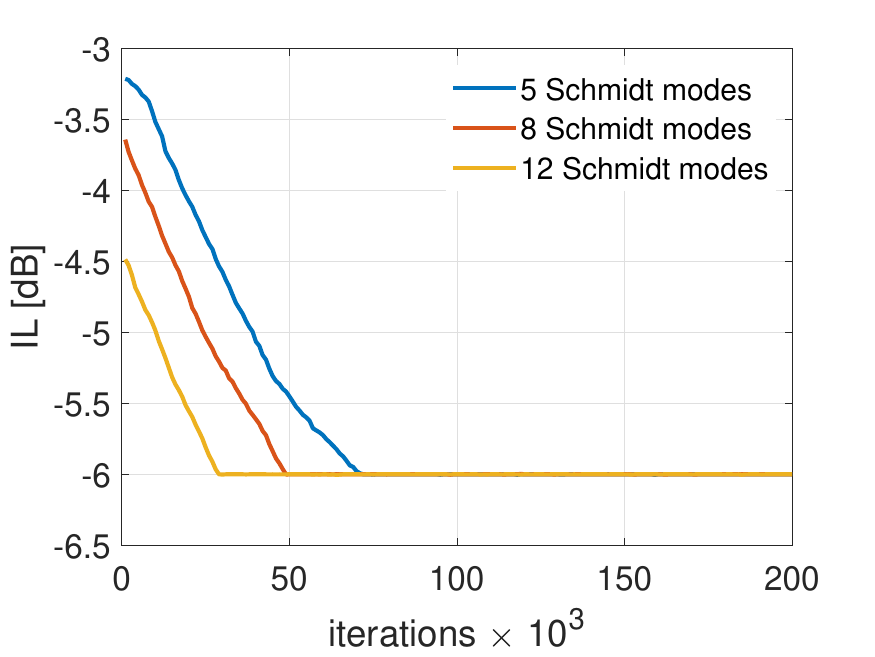}\\
      (d)\includegraphics[width=0.46\textwidth]{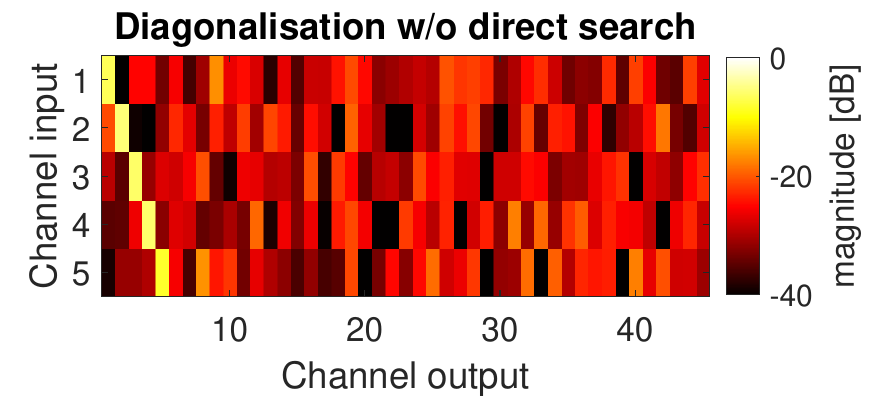}
    \includegraphics[width=0.46\textwidth]{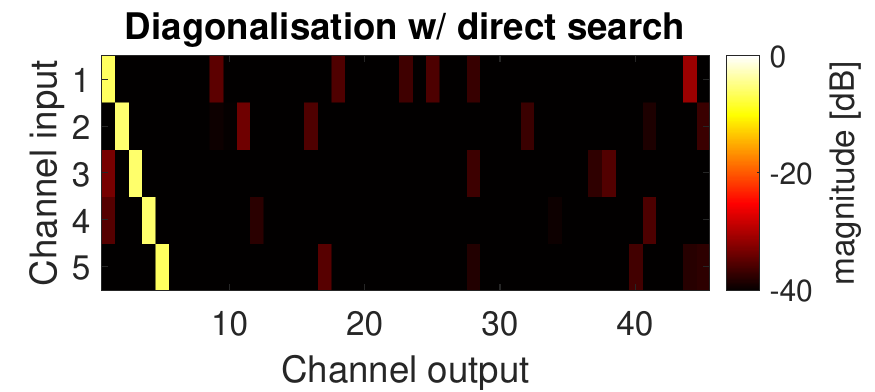}\\
     \caption{Simulation results on Schmidt mode transmission. (a) shows the first 5 Schmidt modes derived from the measured transmission matrix shown in Fig.~\ref{fig:fig1}b. The complex colorbar is shown in the right-hand side. (b) and (c) $\overline{\mathrm{MER}}$ and \gls{il} plotted over iterations using our \gls{ds} method when shaping the first 5, 8 and 12 Schmidt modes at the MMF input, respectively. We use the \gls{wma} result as initial guess. We consider the same 5-passage environment as introduced in Section~\ref{sec:wma} and chose the lower \gls{il} limit at -6~dB. When shaping 5 Schmidt modes using our method, we achieve a gain on $\overline{\mathrm{MER}}$ of $\approx$15~dB in exchange for $\approx$2.8~dB \gls{il}. In the 5 Schmidt mode-case, the $\overline{\mathrm{MER}}$ curve converges after $\approx 500$k iterations. { (d) We simulate channel diagonalisation through an 8~km \gls{mmf} exciting Schmidt modes at the input. We model the transmission by decomposing the shaped field into the LP mode basis and multiply it with the measured transmission matrix. The resulting vector is multiplied with $U^H$ revealing an equalised eigenchannel. The results for transmitting 5 Schmidt modes using \gls{mplc}-based wavefront shaping  at the MMF input are shown in the two panels. Left: Schmidt modes are shaped using only \gls{wma}. Right: Schmidt modes are shaped with our method.}}
    \label{fig:svdResults}
\end{figure}

Considering the 45-mode \gls{mmf} of which we have measured the transmission matrix shown in Fig.~\ref{fig:fig1}b, we can derive spatial transmission eigenchannels by means of \gls{svd} called Schmidt modes~\cite{ho2013linear} $\mathrm{TM}=U \cdot \Sigma \cdot V^H$. 
Launched to an \gls{mmf} with ideal conditions, Schmidt modes enable channel equalisation with zero crosstalk. 
The Schmidt modes are hidden in the columns of $V^H$. 
Each column \textit{i} is filled with $N$ modal coefficients $\gamma_k$ representing the $k$th weight of the corresponding \gls{lp} mode to the \textit{i}th Schmidt mode. 
The field of the \textit{i}th Schmidt mode, e.g. $\Omega_i$, is then calculated by the weighted sum of all \gls{lp} modes $ \Omega_i=\sum^N_k \gamma_{k,i} \Psi_k, $
where $\Psi_j$ indicates the corresponding \gls{lp} mode field. 
Fig.~\ref{fig:svdResults}a shows the first 5 Schmidt modes of the \gls{mmf} under test. 
{
Changing the target basis from \gls{lp} mode to Schmidt modes allows to adapt the \gls{mplc}-based wavefront shaping scheme for a specialised data transmission.
We first shape the Schmidt modes using the standard \gls{wma}.
Next, we use the \gls{wma} result as initial guess and carry out \gls{ds} to reduce cost on $\overline{\mathrm{MER}}$ while not surpassing a lower \gls{il} threshold of $\SI{-6}{\decibel}$.}
The $\overline{\mathrm{MER}}$ standard deviation among the Schmidt modes should be within $\SI{1}{\decibel}$. 
In Fig.~\ref{fig:svdResults}b the $\overline{\mathrm{MER}}$ progression is plotted for varying number of tributaries $M$, while Fig.~\ref{fig:svdResults}c shows the corresponding \gls{il} progression.
With our dynamic cost function design, \gls{ds} optimisation achieves $\SI{15}{\decibel}$ improvement on $\overline{\mathrm{MER}}$ in exchange for $\SI{-2.8}{\decibel}$ \gls{il} deterioration, by shaping a subset of $M=5$ Schmidt modes out of the $N=45$ spatial channels the considered \gls{mmf} supports. 
Increase in the number of tributaries will result in degradation of the achievable improvement on $\overline{\mathrm{MER}}$, if the number of \gls{slm} pixels and the lower \gls{il} limit are kept constant. 
When shaping 8 Schmidt modes in parallel, the achievable average \gls{mer} converges at $\SI{8.5}{\decibel}$, while the $\overline{\mathrm{MER}}$ curve converges at $\SI{4.5}{\decibel}$ for the 12 Schmidt modes-equivalent. 
One particular impediment can be attributed to the fact that becomes more difficult to obtain good starting conditions when increasing the number of tributaries. 
{
As shown in Fig.~\ref{fig:svdResults}c, it can be seen that the initial guesses (i.e. the \gls{wma} results) are deteriorated with a higher \gls{il}, when the number of parallel target Schmidt modes increases.
Consequently, the gain in $\overline{\mathrm{MER}}$ is also hindered to the lower trade-in of \gls{il}, given the lower \gls{il} threshold remains constant.}


{Next, we compare channel diagonalisation via transmitting and receiving Schmidt modes. We shape the singular eigenchannels derived from the transmission matrix at the MMF input using our DS-based optimisation, as explained in the previous paragraph. 
Then we simulate transmission by applying mode decomposition to the shaped fields in the \gls{lp} mode basis matching the basis of our transmission matrix.
We then multiply the mode decomposition result with the measured transmission matrix.
The transmitted signal becomes equalised by multiplying the conjugate transpose output eigenchannel matrix $U^H$. 
Under ideal conditions, this will reveal a diagonal element of $\Sigma$ the power of which scales with the corresponding singular value. 
When shaping the first five Schmidt modes (see Fig.~\ref{fig:svdResults}a) at the MMF input using only the \gls{wma}, we recover the channel equalisation result depicted in the left panel of Fig.~\ref{fig:svdResults}d. In turn, the result shown in the right panel of Fig.~\ref{fig:svdResults}d is found when shaping the Schmidt modes using our \gls{ds}-based optimisation procedure. The crosstalk reduction of $\SI{15}{\decibel}$ on average is quite significant.}

\FloatBarrier

\section{Conclusion}

{
We present a computer-generated hologram optimisation algorithm using direct search to enhance programmable optical data transmission through multimode fibres. Our approach comprises a concatenation of phase modulation planes following a multi-plane light conversion path. While traditional multi-plane light conversion congigurations rely on the wavefront matching algorithm to calculate the phasemasks, we identify unexploited potential in the phasemask design. By using the wavefront matching result as initial state and applying direct search optimisation, we achieve an overall $\SI{15}{\decibel}$ improvement of mode extinction ratio on shaping circularly symmetric linearly polarised modes of the multimode fiber. 
This reward comes at the expense of a targeted deterioration in insertion loss. However, by imposing a lower bound on insertion loss, we avoid entering undesired parameter territory. Similar to the findings reported in Ref.~\cite{kupianskyi2023high}, we observe that relaxing the loss constraints usually yields to higher fidelity, e.g. average mode extinction ratio. }

{
In our simulations, we assume a 5-passage beam transformation with $256\times256$ pixels per passage, amounting to roughly 327k pixels in total. This setup can be directly implemented on a 1M pixel programmable spatial light modulator. Given such configuration, achieving significant improvements, say 10~dB gain in mode extinction ratio, requires at least 250k computational iterations. The convergence efficiency defined as is the ratio of gain to required iterations substantially depends on the selected constraints --- stricter constraints demand more iterations for improvements.}

{
While our method demonstrates clear trade-offs and potential for high-throughput communications, we acknowledge that alternative optimization techniques, such as gradient-based optimisation~\cite{barre2022holographic,kupianskyi2023high}, genetic algorithm~\cite{fickler2020full}, deep learning~\cite{buske2022advanced,mills2022single,buske2024high} where a recent tutorial can be found in Ref.~\cite{hadad2023deep}, or Fourier transform-based optimisation~\cite{chang2015speckle,hayasaki2023automatic}, may provide similar gains with less computational overhead. Nonetheless, this work highlights the additional opportunities available through careful phasemask design optimisation and its impact on mode extinction ratio.}

{
While our direct search optimisation demonstrates significant improvements over the wavefront matching algorithm, it does not guarantee a globally optimal solution, as both methods are limited by the explored search space and computational constraints. The search for a global optimum in the high-dimensional pixel space remains an open challenge, and further research is required to develop algorithms that are capable of reliably converging to such solutions.}


\subsection*{Acknowledgments}
\noindent We thank David~B.~Phillips for insightful discussion. This work was supported by the German Federal Ministry of Education and Research~(BMBF) with the project 6G-life~6KISK001K, the German Research Foundation~(DFG) under grants RO~7348/1-1 and CZ~55/42-2, and the UKRI Future Leaders Fellowship MR/T041218/1.

\subsection*{Disclosures}
The authors declare no conflicts of interest.


\bibliographystyle{bibft}\it
\bibliography{JPhys_template}

\begin{thebibliography}{10}
\expandafter\ifx\csname url\endcsname\relax
  \def\url#1{\texttt{#1}}\fi
\expandafter\ifx\csname urlprefix\endcsname\relax\def\urlprefix{URL }\fi
\expandafter\ifx\csname href\endcsname\relax
  \def\href#1#2{#2} \def\path#1{#1}\fi

\bibitem{essiambre2010capacity}
R.-J. Essiambre, G.~Kramer, P.~J. Winzer, G.~J. Foschini, B.~Goebel, Capacity limits of optical fiber networks, Journal of Lightwave Technology 28~(4) (2010) 662--701.

\bibitem{richardson2010filling}
D.~J. Richardson, Filling the light pipe, Science 330~(6002) (2010) 327--328.

\bibitem{richardson2013space}
D.~J. Richardson, J.~M. Fini, L.~E. Nelson, Space-division multiplexing in optical fibres, Nature photonics 7~(5) (2013) 354--362.

\bibitem{puttnam2021space}
B.~J. Puttnam, G.~Rademacher, R.~S. Lu{\'i}s, Space-division multiplexing for optical fiber communications, Optica 8~(9) (2021) 1186--1203.

\bibitem{arik2016group}
S.~{\"O}. Ar{\i}k, K.-P. Ho, J.~M. Kahn, Group delay management and multiinput multioutput signal processing in mode-division multiplexing systems, Journal of Lightwave Technology 34~(11) (2016) 2867--2880.

\bibitem{ferreira2017semi}
F.~M. Ferreira, C.~S. Costa, S.~Sygletos, A.~D. Ellis, Semi-analytical modelling of linear mode coupling in few-mode fibers, Journal of Lightwave Technology 35~(18) (2017) 4011--4022.

\bibitem{ryf2012mode}
R.~Ryf, S.~Randel, A.~H. Gnauck, C.~Bolle, A.~Sierra, S.~Mumtaz, M.~Esmaeelpour, E.~C. Burrows, R.-J. Essiambre, P.~J. Winzer, et~al., Mode-division multiplexing over 96 km of few-mode fiber using coherent 6$\times$ 6 mimo processing, Journal of Lightwave technology 30~(4) (2012) 521--531.

\bibitem{mori2014few}
T.~Mori, T.~Sakamoto, M.~Wada, T.~Yamamoto, F.~Yamamoto, Few-mode fibers supporting more than two lp modes for mode-division-multiplexed transmission with mimo dsp, Journal of Lightwave Technology 32~(14) (2014) 2468--2479.

\bibitem{ho2011mode}
K.-P. Ho, J.~M. Kahn, Mode-dependent loss and gain: statistics and effect on mode-division multiplexing, Optics express 19~(17) (2011) 16612--16635.

\bibitem{vblast}
G.~J. Foschini, Layered space-time architecture for wireless communication in a fading environment when using multi-element antennas, Bell Labs Technical Journal 1~(2) (1996) 41--59.
\newblock \href {http://dx.doi.org/10.1002/bltj.2015} {\path{doi:10.1002/bltj.2015}}.

\bibitem{cyclic}
K.~Shibahara, T.~Mizuno, H.~Ono, K.~Nakajima, Y.~Miyamoto, Long-haul dmd-unmanaged 6-mode-multiplexed transmission employing cyclic mode-group permutation, in: 2020 Optical Fiber Communications Conference and Exhibition (OFC), 2020, pp. 1--3.

\bibitem{InanComplexity}
B.~Inan, B.~Spinnler, F.~Ferreira, D.~van~den Borne, A.~Lobato, S.~Adhikari, V.~A. J.~M. Sleiffer, M.~Kuschnerov, N.~Hanik, S.~L. Jansen, \href{http://opg.optica.org/oe/abstract.cfm?URI=oe-20-10-10859}{Dsp complexity of mode-division multiplexed receivers}, Opt. Express 20~(10) (2012) 10859--10869.
\newblock \href {http://dx.doi.org/10.1364/OE.20.010859} {\path{doi:10.1364/OE.20.010859}}.
\newline\urlprefix\url{http://opg.optica.org/oe/abstract.cfm?URI=oe-20-10-10859}

\bibitem{Filipe1000}
F.~M. Ferreira, F.~A. Barbosa, Maximizing the capacity of graded-index multimode fibers in the linear regime, Journal of Lightwave Technology (2023) 1--8\href {http://dx.doi.org/10.1109/JLT.2023.3324611} {\path{doi:10.1109/JLT.2023.3324611}}.

\bibitem{popoff2011controlling}
S.~Popoff, G.~Lerosey, M.~Fink, A.~C. Boccara, S.~Gigan, Controlling light through optical disordered media: transmission matrix approach, New Journal of Physics 13~(12) (2011) 123021.

\bibitem{choi2011transmission}
W.~Choi, A.~P. Mosk, Q.-H. Park, W.~Choi, Transmission eigenchannels in a disordered medium, Physical Review B 83~(13) (2011) 134207.

\bibitem{vcivzmar2012exploiting}
T.~{\v{C}}i{\v{z}}m{\'a}r, K.~Dholakia, Exploiting multimode waveguides for pure fibre-based imaging, Nature communications 3~(1) (2012) 1027.

\bibitem{carpenter2016complete}
J.~Carpenter, B.~J. Eggleton, J.~Schr{\"o}der, Complete spatiotemporal characterization and optical transfer matrix inversion of a 420 mode fiber, Optics letters 41~(23) (2016) 5580--5583.

\bibitem{xiong2018complete}
W.~Xiong, C.~W. Hsu, Y.~Bromberg, J.~E. Antonio-Lopez, R.~Amezcua~Correa, H.~Cao, Complete polarization control in multimode fibers with polarization and mode coupling, Light: Science \& Applications 7~(1) (2018) 54.

\bibitem{mazur2019characterization}
M.~Mazur, N.~K. Fontaine, R.~Ryf, H.~Chen, D.~T. Neilson, M.~Bigot-Astruc, F.~Achten, P.~Sillard, A.~Amezcua-Correa, J.~Schr{\"o}der, et~al., Characterization of long multi-mode fiber links using digital holography, in: Optical Fiber Communication Conference, Optical Society of America, 2019, pp. W4C--5.

\bibitem{rothe2019transmission}
S.~Rothe, H.~Radner, N.~Koukourakis, J.~W. Czarske, Transmission matrix measurement of multimode optical fibers by mode-selective excitation using one spatial light modulator, Applied Sciences 9~(1) (2019) 195.

\bibitem{Cheng2023}
S.~Cheng, X.~Zhang, T.~Zhong, H.~Li, H.~Li, L.~Gong, H.~Liu, P.~Lai, \href{https://doi.org/10.1117/1.APN.2.6.066005}{{Nonconvex optimization for optimum retrieval of the transmission matrix of a multimode fiber}}, Advanced Photonics Nexus 2~(6) (2023) 066005.
\newblock \href {http://dx.doi.org/10.1117/1.APN.2.6.066005} {\path{doi:10.1117/1.APN.2.6.066005}}.
\newline\urlprefix\url{https://doi.org/10.1117/1.APN.2.6.066005}

\bibitem{zhong2023}
J.~Zhong, Z.~Wen, Q.~Li, Q.~Deng, Q.~Yang, \href{https://doi.org/10.1117/1.APN.2.5.056007}{{Efficient reference-less transmission matrix retrieval for a multimode fiber using fast Fourier transform}}, Advanced Photonics Nexus 2~(5) (2023) 056007.
\newblock \href {http://dx.doi.org/10.1117/1.APN.2.5.056007} {\path{doi:10.1117/1.APN.2.5.056007}}.
\newline\urlprefix\url{https://doi.org/10.1117/1.APN.2.5.056007}

\bibitem{cao2023controlling}
H.~Cao, T.~{\v{C}}i{\v{z}}m{\'a}r, S.~Turtaev, T.~Tyc, S.~Rotter, Controlling light propagation in multimode fibers for imaging, spectroscopy, and beyond, Advances in Optics and Photonics 15~(2) (2023) 524--612.

\bibitem{leon2010photonic}
S.~G. Leon-Saval, A.~Argyros, J.~Bland-Hawthorn, Photonic lanterns: a study of light propagation in multimode to single-mode converters, Optics Express 18~(8) (2010) 8430--8439.

\bibitem{fontaine2022photonic}
N.~K. Fontaine, J.~Carpenter, S.~Gross, S.~Leon-Saval, Y.~Jung, D.~J. Richardson, R.~Amezcua-Correa, Photonic lanterns, 3-d waveguides, multiplane light conversion, and other components that enable space-division multiplexing, Proceedings of the IEEE 110~(11) (2022) 1821--1834.

\bibitem{alarcon2021few}
A.~Alarcon, J.~Argillander, G.~Lima, G.~B. Xavier, Few-mode-fiber technology fine-tunes losses in quantum communication systems, Physical review applied 16~(3) (2021) 034018.

\bibitem{alarcon2023dynamic}
A.~Alarcon, J.~Argillander, D.~Spegel-Lexne, G.~B. Xavier, Dynamic generation of photonic spatial quantum states with an all-fiber platform, Optics Express 31~(6) (2023) 10673--10683.

\bibitem{velazquez2015six}
A.~Velazquez-Benitez, J.~Alvarado, G.~Lopez-Galmiche, J.~Antonio-Lopez, J.~Hern{\'a}ndez-Cordero, J.~Sanchez-Mondragon, P.~Sillard, C.~Okonkwo, R.~Amezcua-Correa, Six mode selective fiber optic spatial multiplexer, Optics letters 40~(8) (2015) 1663--1666.

\bibitem{Fontaine12}
N.~K. Fontaine, R.~Ryf, J.~Bland-Hawthorn, S.~G. Leon-Saval, \href{https://opg.optica.org/oe/abstract.cfm?URI=oe-20-24-27123}{Geometric requirements for photonic lanterns in space division multiplexing}, Opt. Express 20~(24) (2012) 27123--27132.
\newblock \href {http://dx.doi.org/10.1364/OE.20.027123} {\path{doi:10.1364/OE.20.027123}}.
\newline\urlprefix\url{https://opg.optica.org/oe/abstract.cfm?URI=oe-20-24-27123}

\bibitem{Velazquez18}
A.~Velázquez-Benítez, J.~Antonio-López, J.~alvarado zacarias, N.~Fontaine, R.~Ryf, H.~Chen, J.~Hernández-Cordero, P.~Sillard, C.~Okonkwo, S.~Leon-Saval, R.~Amezcua-Correa, Scaling photonic lanterns for space-division multiplexing, Scientific Reports 8.
\newblock \href {http://dx.doi.org/10.1038/s41598-018-27072-2} {\path{doi:10.1038/s41598-018-27072-2}}.

\bibitem{vellekoop2015feedback}
I.~M. Vellekoop, Feedback-based wavefront shaping, Optics express 23~(9) (2015) 12189--12206.

\bibitem{cao2022shaping}
H.~Cao, A.~P. Mosk, S.~Rotter, Shaping the propagation of light in complex media, Nature Physics 18~(9) (2022) 994--1007.

\bibitem{gomes2022near}
A.~D. Gomes, S.~Turtaev, Y.~Du, T.~{\v{C}}i{\v{z}}m{\'a}r, Near perfect focusing through multimode fibres, Optics Express 30~(7) (2022) 10645--10663.

\bibitem{ho2013linear}
K.-P. Ho, J.~M. Kahn, Linear propagation effects in mode-division multiplexing systems, Journal of lightwave technology 32~(4) (2013) 614--628.

\bibitem{mosk2012controlling}
A.~P. Mosk, A.~Lagendijk, G.~Lerosey, M.~Fink, Controlling waves in space and time for imaging and focusing in complex media, Nature photonics 6~(5) (2012) 283--292.

\bibitem{kim2012maximal}
M.~Kim, Y.~Choi, C.~Yoon, W.~Choi, J.~Kim, Q.-H. Park, W.~Choi, Maximal energy transport through disordered media with the implementation of transmission eigenchannels, Nature photonics 6~(9) (2012) 581--585.

\bibitem{sarma2016control}
R.~Sarma, A.~G. Yamilov, S.~Petrenko, Y.~Bromberg, H.~Cao, Control of energy density inside a disordered medium by coupling to open or closed channels, Physical review letters 117~(8) (2016) 086803.

\bibitem{yilmaz2019transverse}
H.~Y{\i}lmaz, C.~W. Hsu, A.~Yamilov, H.~Cao, Transverse localization of transmission eigenchannels, Nature Photonics 13~(5) (2019) 352--358.

\bibitem{fabioPMs}
F.~A. Barbosa, F.~M. Ferreira, Scaling spatial multiplexing with principal modes, in: 2022 IEEE Photonics Conference (IPC), IEEE, 2022, pp. 1--2.

\bibitem{carpenter2015observation}
J.~Carpenter, B.~J. Eggleton, J.~Schr{\"o}der, Observation of eisenbud--wigner--smith states as principal modes in multimode fibre, Nature Photonics 9~(11) (2015) 751--757.

\bibitem{fabioPMsJournal}
F.~A. Barbosa, F.~M. Ferreira, On a scalable path for multimode sdm transmission, Journal of Lightwave Technology (2023) 1--10\href {http://dx.doi.org/10.1109/JLT.2023.3308777} {\path{doi:10.1109/JLT.2023.3308777}}.

\bibitem{rothe2023securing}
S.~Rothe, K.-L. Besser, D.~Krause, R.~Kuschmierz, N.~Koukourakis, E.~Jorswieck, J.~W. Czarske, \href{https://spj.science.org/doi/abs/10.34133/research.0065}{Securing data in multimode fibers by exploiting mode-dependent light propagation effects}, Research 6 (2023) 0065.
\newblock \href {http://dx.doi.org/10.34133/research.0065} {\path{doi:10.34133/research.0065}}.
\newline\urlprefix\url{https://spj.science.org/doi/abs/10.34133/research.0065}

\bibitem{labroille2014efficient}
G.~Labroille, B.~Denolle, P.~Jian, P.~Genevaux, N.~Treps, J.-F. Morizur, Efficient and mode selective spatial mode multiplexer based on multi-plane light conversion, Optics express 22~(13) (2014) 15599--15607.

\bibitem{fontaine2017design}
N.~K. Fontaine, R.~Ryf, H.~Chen, D.~Neilson, J.~Carpenter, Design of high order mode-multiplexers using multiplane light conversion, in: 2017 European Conference on Optical Communication (ECOC), IEEE, 2017, pp. 1--3.

\bibitem{fontaine2019laguerre}
N.~K. Fontaine, R.~Ryf, H.~Chen, D.~T. Neilson, K.~Kim, J.~Carpenter, Laguerre-gaussian mode sorter, Nature communications 10~(1) (2019) 1--7.

\bibitem{mounaix2020time}
M.~Mounaix, N.~K. Fontaine, D.~T. Neilson, R.~Ryf, H.~Chen, J.~C. Alvarado-Zacarias, J.~Carpenter, Time reversed optical waves by arbitrary vector spatiotemporal field generation, Nature communications 11~(1) (2020) 1--7.

\bibitem{fontaine2021hermite}
N.~K. Fontaine, H.~Chen, M.~Mazur, L.~Dallachiesa, K.~Kim, R.~Ryf, D.~Neilson, J.~Carpenter, Hermite-gaussian mode multiplexer supporting 1035 modes, in: Optical Fiber Communication Conference, Optica Publishing Group, 2021, pp. M3D--4.

\bibitem{rademacher2021peta}
G.~Rademacher, B.~J. Puttnam, R.~S. Lu{\'i}s, T.~A. Eriksson, N.~K. Fontaine, M.~Mazur, H.~Chen, R.~Ryf, D.~T. Neilson, P.~Sillard, et~al., Peta-bit-per-second optical communications system using a standard cladding diameter 15-mode fiber, Nature Communications 12~(1) (2021) 1--7.

\bibitem{wen2021scalable}
H.~Wen, Y.~Zhang, R.~Sampson, N.~K. Fontaine, N.~Wang, S.~Fan, G.~Li, Scalable non-mode selective hermite--gaussian mode multiplexer based on multi-plane light conversion, Photonics Research 9~(2) (2021) 88--97.

\bibitem{kupianskyi2023high}
H.~Kupianskyi, S.~A. Horsley, D.~B. Phillips, High-dimensional spatial mode sorting and optical circuit design using multi-plane light conversion, APL Photonics 8~(2).

\bibitem{pohle2023intelligent}
D.~Pohle, F.~A. Barbosa, F.~M. Ferreira, J.~Czarske, S.~Rothe, Intelligent self calibration tool for adaptive few-mode fiber multiplexers using multiplane light conversion, Journal of the European Optical Society-Rapid Publications 19~(1).

\bibitem{billaud2019optimal}
A.~Billaud, F.~Gomez, D.~Allioux, N.~Laurenchet, P.~Jian, O.~Pinel, G.~Labroille, Optimal coherent beam combining based on multi-plane light conversion for high throughput optical feeder links, in: 2019 IEEE International Conference on Space Optical Systems and Applications (ICSOS), IEEE, 2019, pp. 1--5.

\bibitem{kupianskyi2024all}
H.~Kupianskyi, S.~A. Horsley, D.~B. Phillips, All-optically untangling light propagation through multimode fibers, Optica 11~(1) (2024) 101--112.

\bibitem{fickler2020full}
R.~Fickler, F.~Bouchard, E.~Giese, V.~Grillo, G.~Leuchs, E.~Karimi, Full-field mode sorter using two optimized phase transformations for high-dimensional quantum cryptography, Journal of Optics 22~(2) (2020) 024001.

\bibitem{butaite2022build}
U.~G. B{\=u}tait{\.e}, H.~Kupianskyi, T.~{\v{C}}i{\v{z}}m{\'a}r, D.~B. Phillips, How to build the “optical inverse” of a multimode fibre, Intelligent Computing.

\bibitem{lin2018all}
X.~Lin, Y.~Rivenson, N.~T. Yardimci, M.~Veli, Y.~Luo, M.~Jarrahi, A.~Ozcan, All-optical machine learning using diffractive deep neural networks, Science 361~(6406) (2018) 1004--1008.

\bibitem{zhou2022photonic}
H.~Zhou, J.~Dong, J.~Cheng, W.~Dong, C.~Huang, Y.~Shen, Q.~Zhang, M.~Gu, C.~Qian, H.~Chen, et~al., Photonic matrix multiplication lights up photonic accelerator and beyond, Light: Science \& Applications 11~(1) (2022) 1--21.

\bibitem{yildirim2024nonlinear}
M.~Yildirim, N.~U. Dinc, I.~Oguz, D.~Psaltis, C.~Moser, Nonlinear processing with linear optics, Nature Photonics (2024) 1--7.

\bibitem{sakamaki2007new}
Y.~Sakamaki, T.~Saida, T.~Hashimoto, H.~Takahashi, New optical waveguide design based on wavefront matching method, Journal of lightwave technology 25~(11) (2007) 3511--3518.

\bibitem{tsang2018review}
P.~Tsang, T.-C. Poon, Y.~Wu, Review of fast methods for point-based computer-generated holography, Photonics Research 6~(9) (2018) 837--846.

\bibitem{rothe2021benchmarking}
S.~Rothe, P.~Daferner, S.~Heide, D.~Krause, F.~Schmieder, N.~Koukourakis, J.~W. Czarske, Benchmarking analysis of computer generated holograms for complex wavefront shaping using pixelated phase modulators, Optics Express 29~(23) (2021) 37602--37616.

\bibitem{fontaine2019laguerreSupp}
N.~K. Fontaine, R.~Ryf, H.~Chen, D.~T. Neilson, K.~Kim, J.~Carpenter, Laguerre-gaussian mode sorter supplemental information.

\bibitem{gloge1971weakly}
D.~Gloge, Weakly guiding fibers, Applied optics 10~(10) (1971) 2252--2258.

\bibitem{snyder1983optical}
A.~Snyder, Optical waveguide theory (1983).

\bibitem{siegman1973hermite}
A.~E. Siegman, Hermite--gaussian functions of complex argument as optical-beam eigenfunctions, JOSA 63~(9) (1973) 1093--1094.

\bibitem{seldowitz1987synthesis}
M.~A. Seldowitz, J.~P. Allebach, D.~W. Sweeney, Synthesis of digital holograms by direct binary search, Applied optics 26~(14) (1987) 2788--2798.

\bibitem{clark1996direct}
M.~Clark, R.~Smith, A direct-search method for the computer design of holograms, Optics Communications 124~(1-2) (1996) 150--164.

\bibitem{christopher2020holographic}
P.~J. Christopher, R.~Mouthaan, G.~S. Gordon, T.~D. Wilkinson, Holographic predictive search: Extending the scope, Optics Communications 467 (2020) 125701.

\bibitem{RotheGithub}
S.~Rothe, F.~M. Ferreira, \href{https://github.com/StefRot1992/direct-search-mode-multiplexer}{Direct search optimisation algorithm for mode multiplexers} (2024).
\newline\urlprefix\url{https://github.com/StefRot1992/direct-search-mode-multiplexer}

\bibitem{barre2022holographic}
N.~Barr{\'e}, A.~Jesacher, Holographic beam shaping of partially coherent light, Optics Letters 47~(2) (2022) 425--428.

\bibitem{buske2022advanced}
P.~Buske, A.~V{\"o}ll, M.~Eisebitt, J.~Stollenwerk, C.~Holly, Advanced beam shaping for laser materials processing based on diffractive neural networks, Optics express 30~(13) (2022) 22798--22816.

\bibitem{mills2022single}
B.~Mills, J.~A. Grant-Jacob, M.~Praeger, R.~W. Eason, J.~Nilsson, M.~N. Zervas, Single step phase optimisation for coherent beam combination using deep learning, Scientific Reports 12~(1) (2022) 5188.

\bibitem{buske2024high}
P.~Buske, O.~Hofmann, A.~Bonnhoff, J.~Stollenwerk, C.~Holly, High fidelity laser beam shaping using liquid crystal on silicon spatial light modulators as diffractive neural networks, Optics Express 32~(5) (2024) 7064--7078.

\bibitem{hadad2023deep}
B.~Hadad, S.~Froim, E.~Yosef, R.~Giryes, A.~Bahabad, Deep learning in optics-a tutorial, Journal of Optics.

\bibitem{chang2015speckle}
C.~Chang, J.~Xia, L.~Yang, W.~Lei, Z.~Yang, J.~Chen, Speckle-suppressed phase-only holographic three-dimensional display based on double-constraint gerchberg--saxton algorithm, Applied optics 54~(23) (2015) 6994--7001.

\bibitem{hayasaki2023automatic}
Y.~Hayasaki, R.~Onodeara, K.~Kumagai, S.~Hasegawa, Automatic generation of a holographically shaped beam in an actual optical system for use in material laser processing, Optics Express 31~(2) (2023) 1982--1991.

\end{thebibliography}






\end{document}